\documentclass[useAMS,usenatbib]{mn2e}
\usepackage{graphicx}
\usepackage{psfig}

\title[Condensation temperature trends]{Condensation temperature trends among stars with planets}
\author[G. Gonzalez]{Guillermo Gonzalez$^{1}$\thanks{E-mail:
gonzog@iastate.edu} \\
$^{1}$Iowa State University, Department of Physics and Astronomy, Ames, 
IA 50011}

\begin{document}

\date{Accepted December 5. Received November 30; in original form November 3}

\pagerange{\pageref{firstpage}--\pageref{lastpage}} \pubyear{??}

\maketitle

\label{firstpage}

\begin{abstract}
Results from detailed spectroscopic analyses of stars hosting massive planets are employed to search for trends between abundances and condensation temperatures. The elements C, S, Na, Mg, Al,  Ca, Sc, Ti, V, Cr, Mn, Fe, Ni and Zn are included in the analysis of 64 stars with planets and 33 comparison stars. No significant trends are evident in the data. This null result suggests that accretion of rocky material onto the photospheres of stars with planets is not the primary explanation for their high metallicities. However, the differences between the solar photospheric and meteoritic abundances do display a weak but significant trend with condensation temperature. This suggests that the metallicity of the sun's envelope has been enriched relative to its interior by about 0.07 dex.
\end{abstract}


\section{Introduction}

Spectroscopic analyses of stars with planets (SWPs) have revealed that they, as a group, exhibit higher metallicity than randomly selected samples of nearby stars \citep{l03, s04, FV05}. \citet{g03} reviewed the three explanations proposed to account for this difference. One of these, the "self enrichment" hypothesis, posits that SWPs accreted sufficient mass of metals to enhance the metallicity of their photospheres by observable amounts.

\citet{g97} suggested that accretion of metals by a star might be seen via its effects on the correlation between its photospheric abundances and the elemental condensation temperatures (T$_{\rm c}$). This would be expected if, for example, a star accretes rocky material that formed significantly inside the water ice radius in the protoplanetary disk. Such material would be relatively more depleted in volatiles (e.g., C, N and O) than in refractories (e.g., Al, Ca and Ti). Material in the inner disk could find its way into the central star, for example, from gravitational scattering by a migrating giant planet.

\citet{s01} were the first to search for anomalous trends between abundances and T$_{\rm c}$ among SWPs. They compared the slopes from some 15 elements in 30 SWPs and 102 comparison stars. They found 5 or 6 SWPs with significantly higher abundance-T$_{\rm c}$ slopes than the comparison stars. \citet{g03} also noted that these stars tended to be hotter. However, \citet{t01} and \citet{s02} analyzed 14 and 12 SWPs, respectively, finding no significant correlations between abundances and T$_{\rm c}$. 

\citet{gr01} studied six common proper motion main sequence stars. For one pair, HD 219542 A and B, they found a significant trend between the differences in their abundances and T$_{\rm c}$. \citet{s03} also studied the pair, but they failed to find such a trend.

Taken together, these studies leave the question of abundance-T$_{\rm c}$ trends unresolved. Over the last two years, many more SWPs have been discovered and analyzed using detailed spectroscopic methods. In this study we revisit the question of abundance-T$_{\rm c}$ trends among SWPs using the most recent data sets. We discuss the sample preparation, analysis and interpretation of results. We also search for a trend between the solar photospheric and meteoritic abundance differences and T$_{\rm c}$.

\section{Sample preparation}

A weakness of the Smith et al. study was the heterogeneous nature of their abundance data. In order to avoid possible systematic offsets among different sets of abundance determinations and between the SWP and comparison stars samples, the best strategy is to employ abundance data from a single source. With this in mind, we have selected for our analysis published data for SWPs and comparison stars from the following studies: C, S, Zn, Cu \citep{e04}; Na, Mg, Al \citep{b05}; Si, Ca, Sc, Ti, V, Cr, Mn, Ni \citep{b03}. These studies make use of the same analysis methods. The atmospheric parameters for the SWPs and comparison stars are taken from \citet{s04}. The data culled from these studies formed the first version of our database. It contains abundance\footnotetext{We remind the reader that we are here employing elemental abundance determinations relative to those in the sun.} data for 76 SWPs and 37 comparison stars. More elements could have been included, but the number of stars lacking a complete set of abundance determinations increases as the number of elements is increased.

Still, some of the stars in the database lack abundance determinations for the full suite of elements listed above. To mitigate this deficit, we also included abundance determinations from \citet{gl00}, \citet{g01} and \citet{g05} for the following elements: C, S, Na, Al and Mn. The offsets between the abundances reported in the two sets of studies for each element were calculated using stars common to both studies and then applied prior to importing these data into the database. Typically, these other abundance determinations allowed us to fill in missing data for two or three SWPs per element.

Even with these additions, the database still contains missing data; 64 SWPs and 33 comparison stars in the database are missing at most one element abundance value. Since it is important that the same elements be analyzed for the SWPs and comparison stars, only these 97 stars were included in the final version of the database.

One of the analyses described in the next section make use of the uncertainties in the abundance determinations. We adopted the uncertainties reported in each respective study. In a few instances uncertainties were reported as "0" (mostly for Mn) in the original studies; in such cases, an average value of the uncertainty determined from other stars in the study was adopted instead.

Values for T$_{\rm c}$ for each element are from Table 8 of \citet{lod03}. The T$_{\rm c}$ value for a given element corresponds to the temperature at which 50\% of it condenses out of the gas phase as the temperature is reduced. The values of T$_{\rm c}$ for the elements included in the database range from 40 K for C to 1659 K for Sc; the C abundance, in particular, provides considerable leverage in calculating the slope between abundance and T$_{\rm c}$ for each star.

\section{Analysis}
\subsection{Stars with Planets}

For each star in the database, we determined a linear least squares solution between the abundances and T$_{\rm c}$ values. Two versions of least squares analysis were employed. The first makes use of weighted least squares using the uncertainties in the abundances. The results of this analysis are shown in Fig.~1 with the T$_{\rm c}$ slope values plotted against [Fe/H]. Four stars, all SWPs, deviate significantly from the general trend; HD 114783 is much higher than all the other points, while HD 19994, HD 9826 and HD 179949 are slightly low.

\begin{figure}
 \includegraphics[width=3.3in]{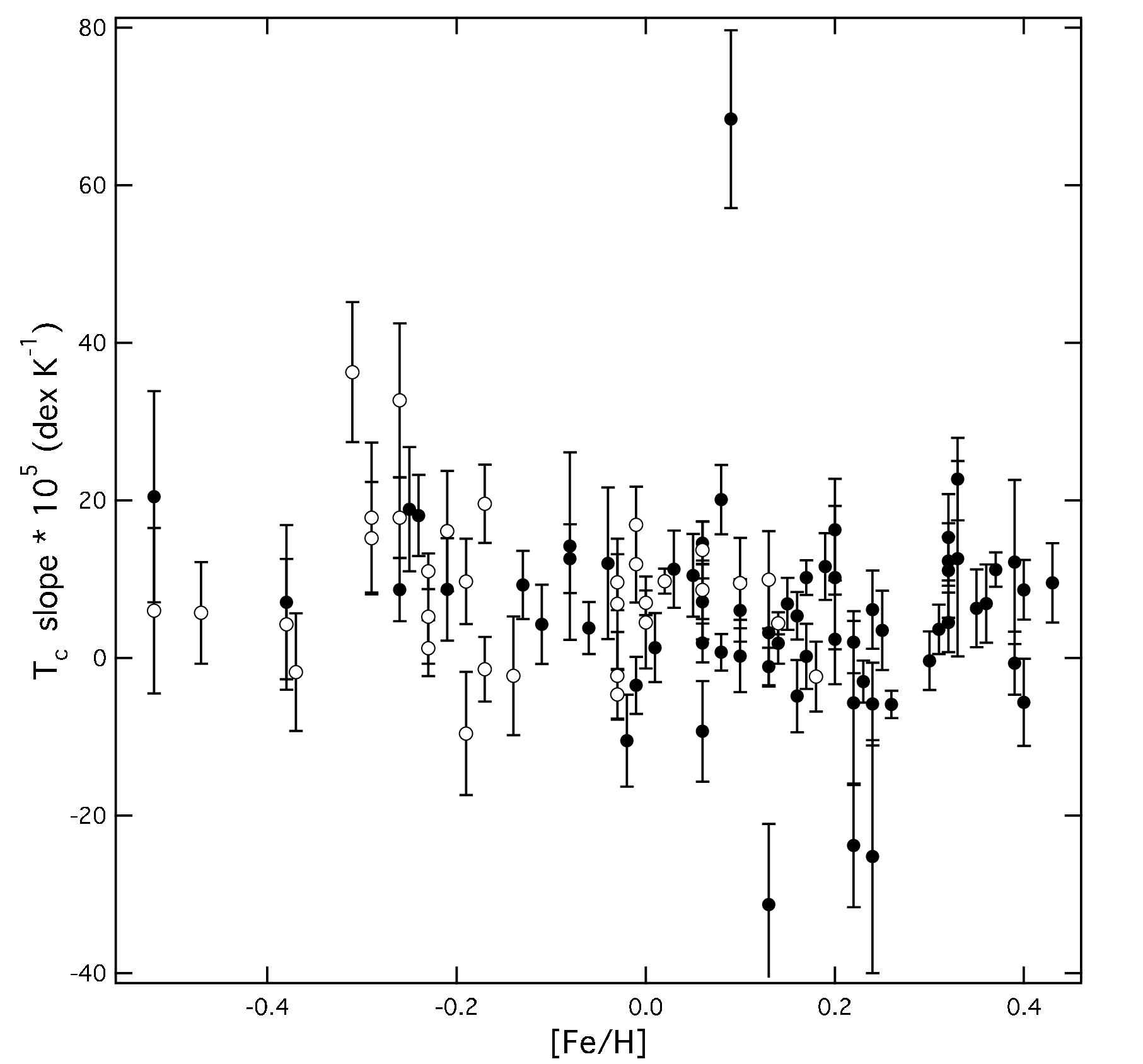}
 \caption{T$_{\rm c}$ slope is plotted against [Fe/H] for SWPs (dots) and comparison stars (open circles) The T$_{\rm c}$ slope for each star was determined from a weighted least squares analysis.}
\end{figure}

The second method is a standard unweighted linear least squares analysis. We show the results of this analysis in Fig.~2. In this case only one star, HD 5133 -- a comparison star, deviates significantly from the others. The overall scatter among the points in Fig.~2 is also smaller than in Fig.~1. While in principle weighted least squares should be preferable to unweighted least squares, in this case it is not.

\begin{figure}
  \includegraphics[width=3.3in]{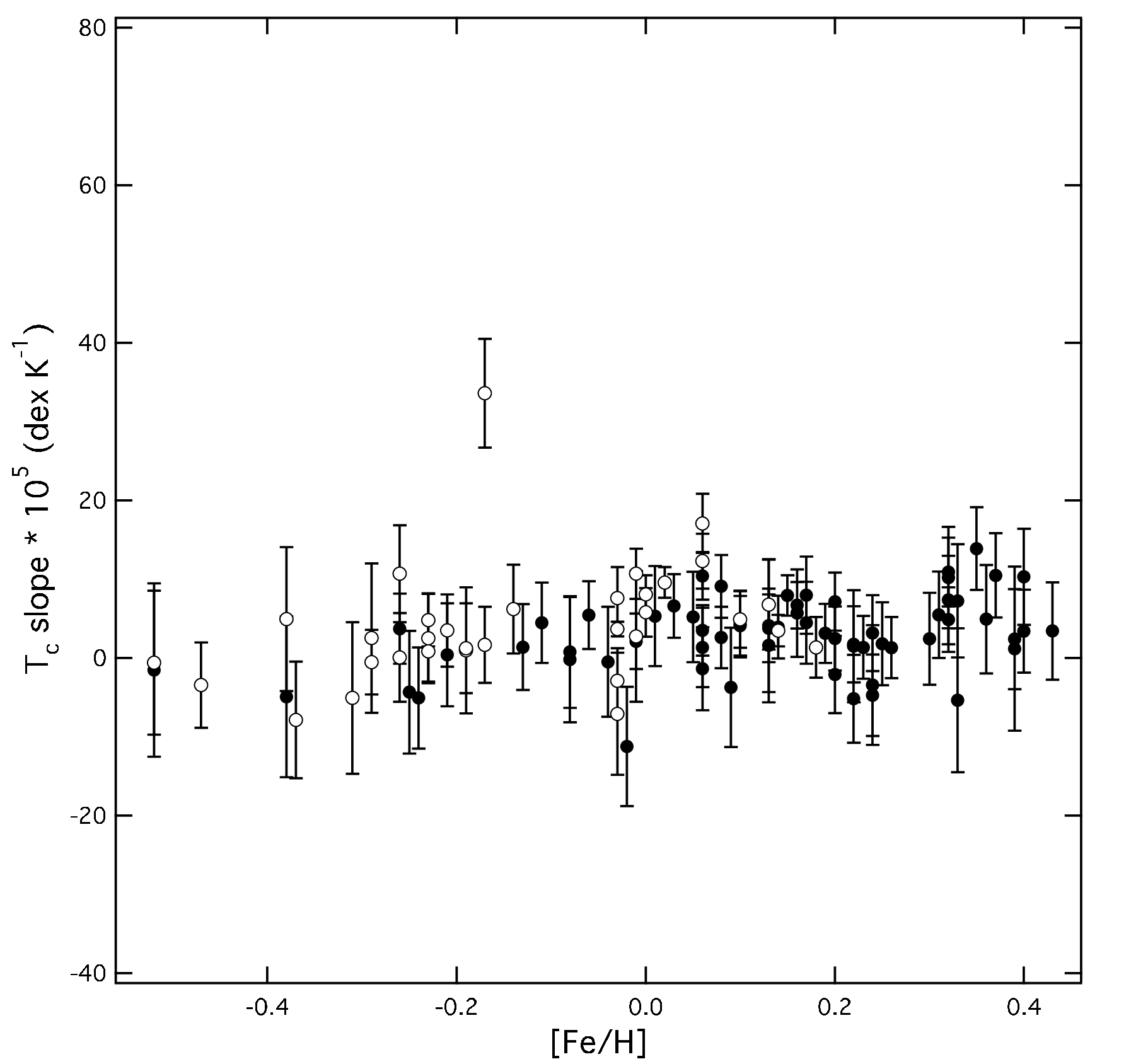}
 \caption{Same as Fig.~1 but using standard unweighted least squares analysis.}
\end{figure}

In order to obtain reliable results from weighted least squares, it is necessary to know the uncertainties accurately. Those elements for which the uncertainties in the abundances are underestimated will be given too much weight in the solution. To illustrate this, we plot abundances against T$_{\rm c}$ for HD 9826 and HD 114783 in Fig.~3. From the weighted least squares analysis, the slope for HD 9826 was found to be negative, while it was positive for HD 114783. In each case, the derived slope is strongly biased by 2 or 3 data points that have much smaller uncertainties than the other data (see Fig.~3). However, for both stars the slopes derived form the standard least squares analysis yielded values indistinguishable from zero. For the remainder of the analysis we will rely on the unweighted analysis results.

\begin{figure}
  \includegraphics[width=3.3in]{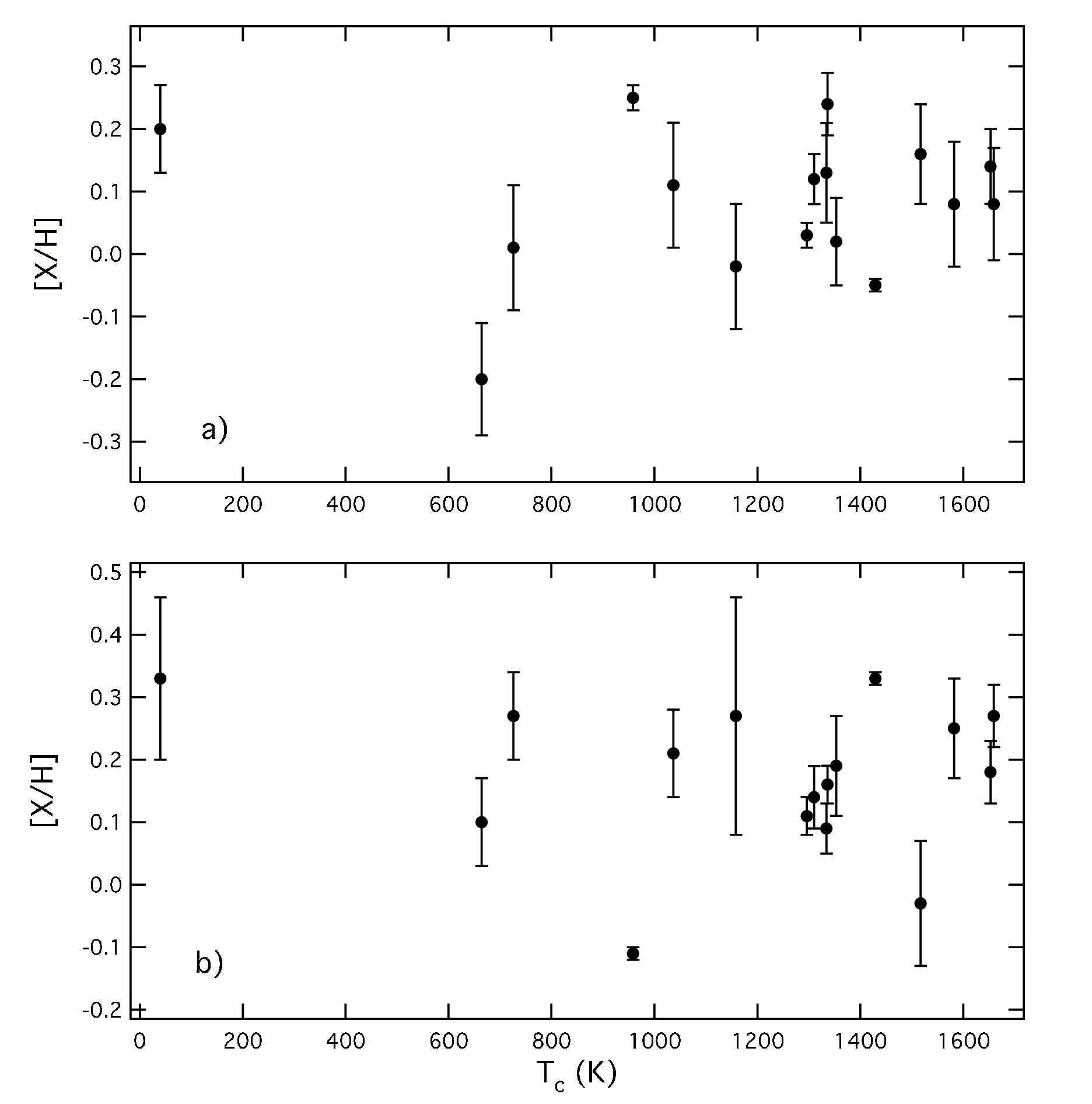}
 \caption{Abundances plotted against T$_{\rm c}$ for HD 9826 (panel a) and HD 114783 (panel b).}
\end{figure}

As noted by \citet{g03}, a trend between the T$_{\rm c}$ slope and T$_{\rm eff}$ can be a signature of accretion. Hotter stars have thinner outer convection zones and are thus more likely to display the effects of accretion. To test for this, we performed a multiple linear regression analysis on the unweighted T$_{\rm c}$ slopes with effective temperature (T$_{\rm eff}$) and [Fe/H] as independent parameters. The resulting equation for the T$_{\rm c}$ slopes of the SWPs is:

$-(1.7 \pm 0.8)\times 10^{-4} + (3 \pm 1)\times 10^{-8}$ T$_{\rm eff}  + (0.9 \pm 0.3)\times 10^{\rm -4} {\rm [Fe/H]}$

and the equation for the comparison stars is (excluding the outlier):

$(1.3 \pm 1.4)\times 10^{-4} - (1 \pm 2)\times 10^{-8}$ T$_{\rm eff} + (1.5 \pm 0.5)\times 10^{\rm -4} {\rm [Fe/H]}$

The zero-points and T$_{\rm eff}$ coefficients differ marginally at the one-$\sigma$ level. We show the two fits in Fig.~4. 

\begin{figure}
 \includegraphics[width=3.3in]{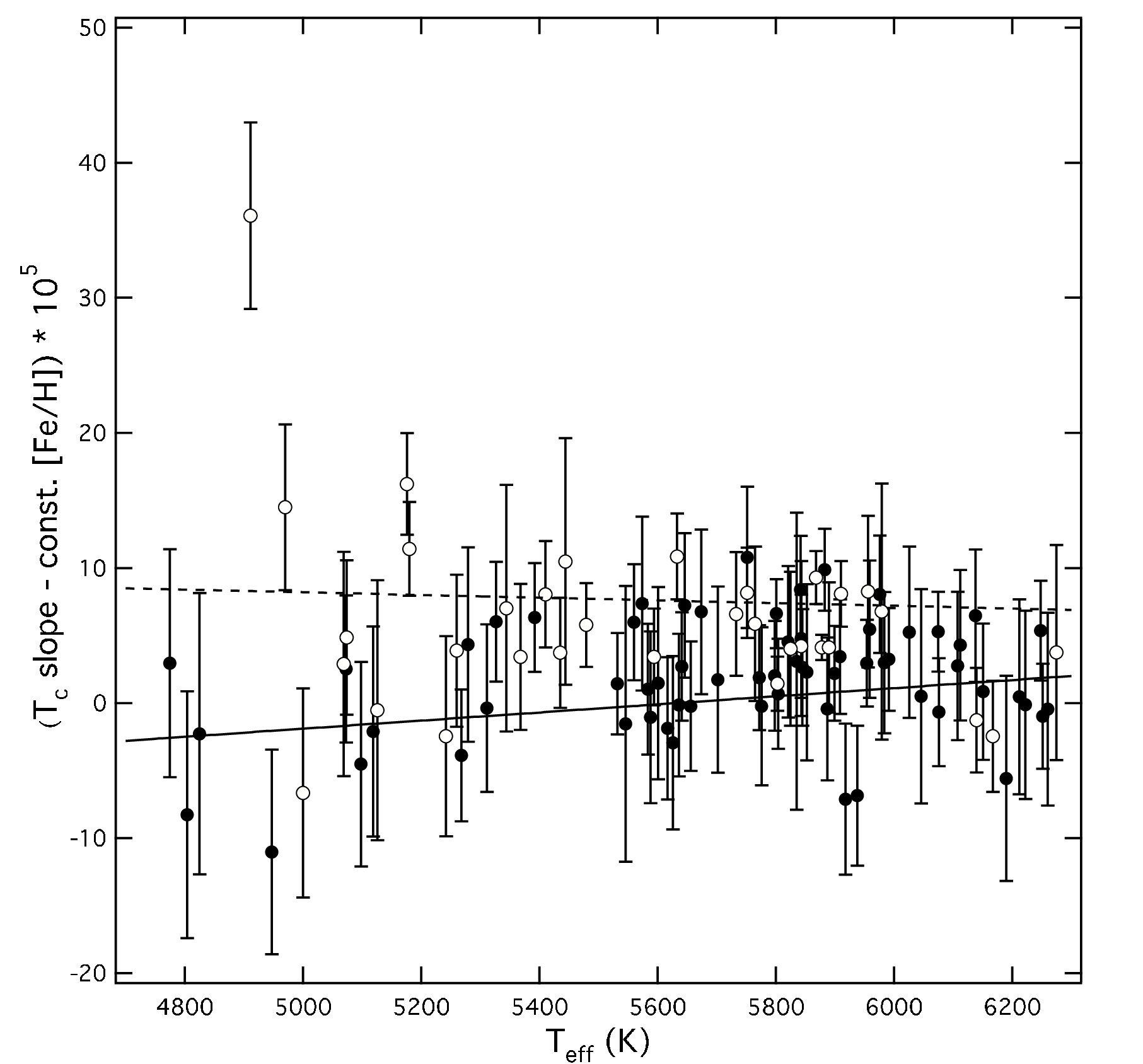}
 \caption{T$_{\rm c}$ slope corrected for [Fe/H] trends plotted against effective temperature. The solid line is a least squares fit to the SWPs, and the dashed line is a fit to the comparison stars.}
\end{figure}

\subsection{Solar System Abundances}

\citet{g97,g03} noted that the signature of accretion of high T$_{\rm c}$ material in the sun might be detectable by comparing the solar photospheric abundances to those of the CI carbonaceous chondrite meteorites (hereafter, chondrites). The elemental abundances in chondrites closely track those in the solar photosphere. Indeed, this is often cited as strong evidence for: 1) the common origin of the sun and meteorites and 2) the primitive nature of the chondrites. However, the chondrites are deficient relative to the solar photosphere in the most volatile elements, H, He, C, N and O. 

If the sun has accreted rocky material, then its atmosphere should have been enriched in elements with high T$_{\rm c}$ values. The differences between the photospheric and chondritic abundances should display an increasing trend with T$_{\rm c}$. 

To test for a trend with T$_{\rm c}$ in the sun, we calculated the differences between the solar photospheric and chondritic abundances tabulated in two recent studies: Table 1 of \citet{lod03} and Table 1 of \citet{asp05}. We selected elements from these studies with photospheric and chondritic abundance uncertainties less than or equal to 0.15 dex; the only exceptions to this rule are Li, which we excluded since it is known to have been depleted in the sun's envelope via proton capture over its lifetime, and Be since it was possibly slightly depleted by the same process. The uncertainties in the abundance differences were calculated by summing the photospheric and chondritic abundances in quadrature. The resulting abundance differences are shown in Figs.~5 and 6.

\begin{figure}
 \includegraphics[width=3.3in]{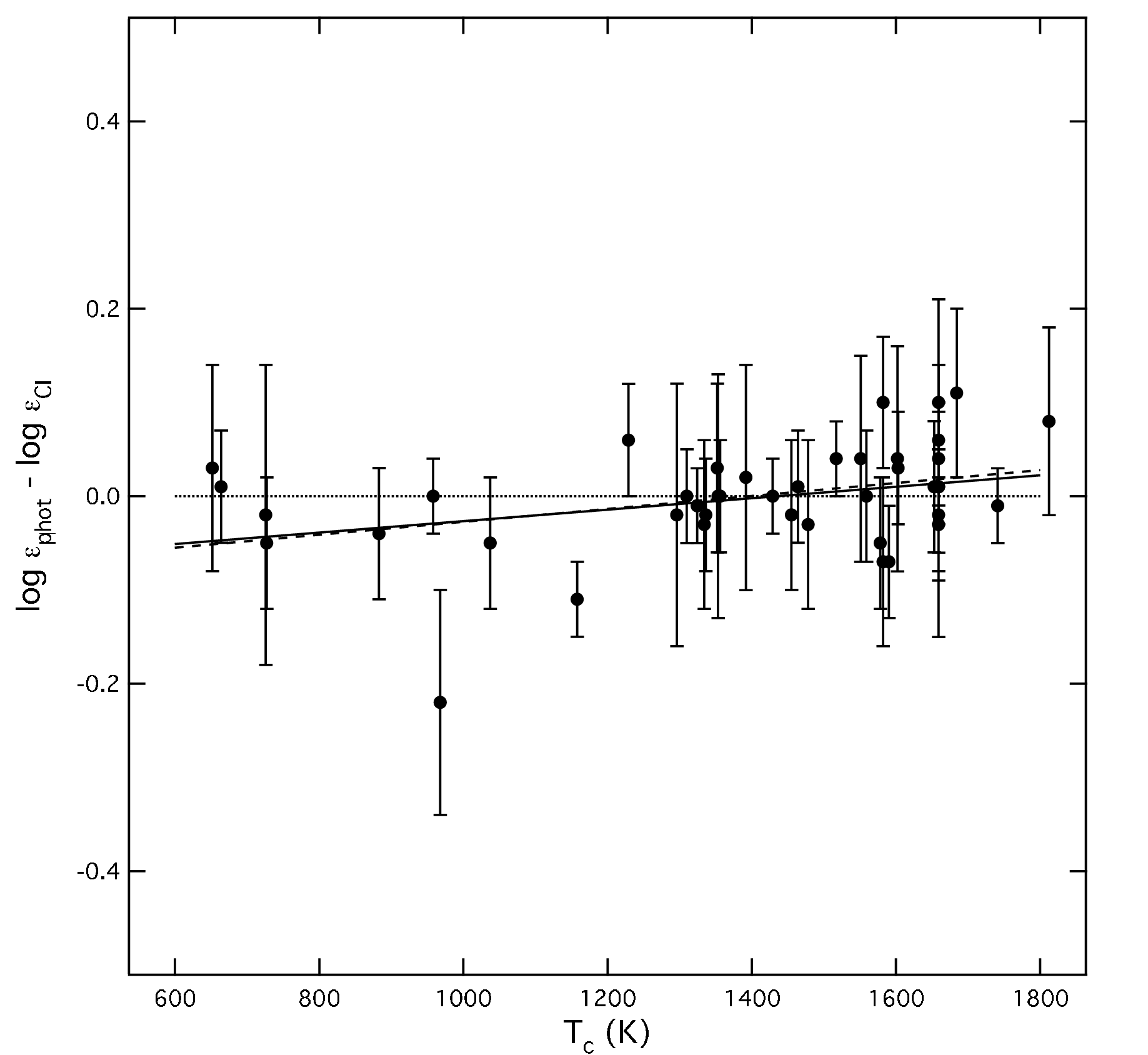}
 \caption{Differences in photospheric and chondritic abundances plotted against T$_{\rm c}$ for 43 elements from Table 1 of \citet{lod03}. The weighted least squares fit is shown as a solid line, and the unweighted fit is shown as a dashed line. The dotted line corresponds to zero difference.}
\end{figure}

\begin{figure}
 \includegraphics[width=3.3in]{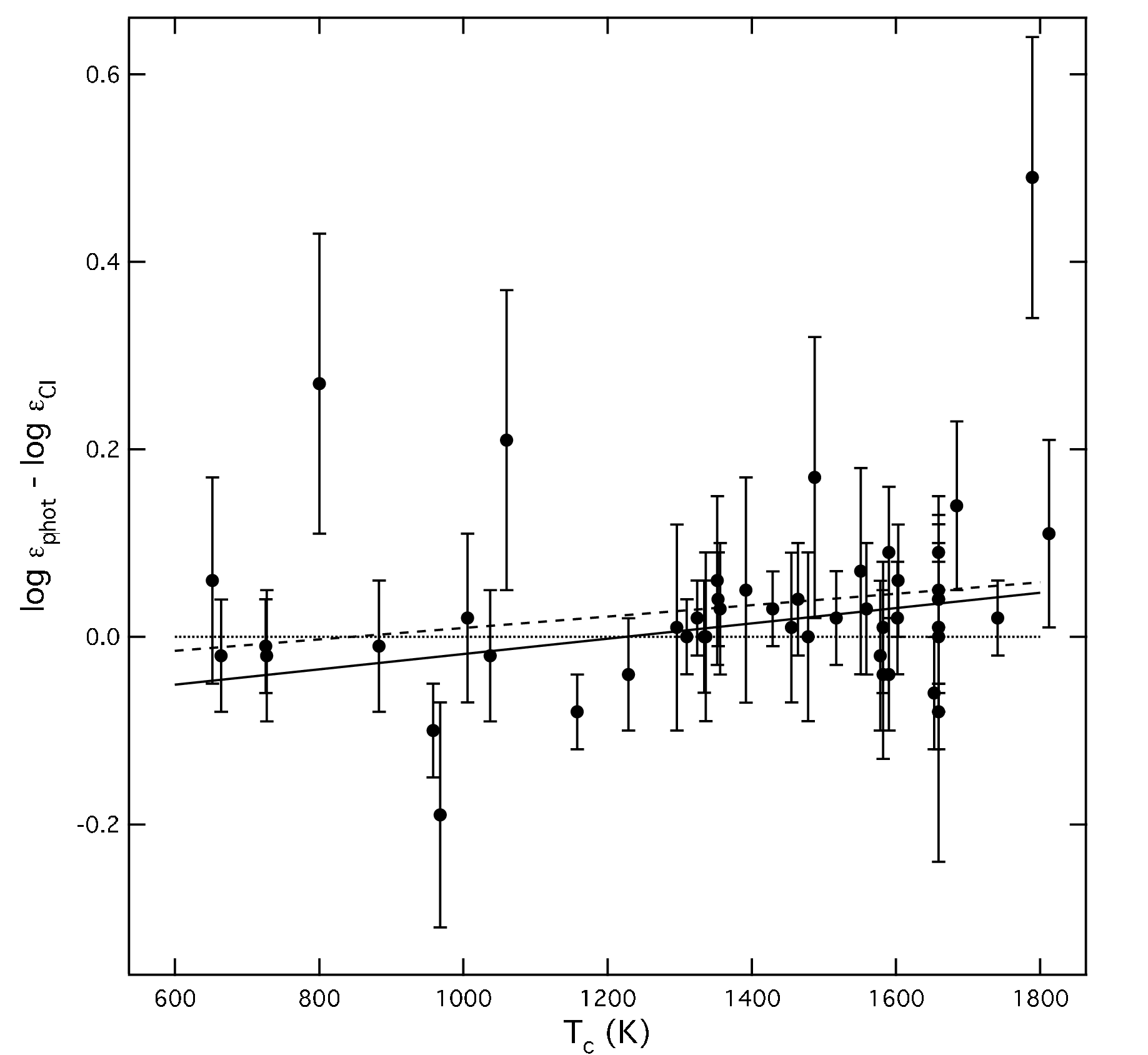}
 \caption{Differences in photospheric and chondritic abundances plotted against T$_{\rm c}$ for 50 elements from Table 1 of \citet{asp05}. All else as in Fig.~5.}
\end{figure}

The weighted and unweighted least squares fits shown in Fig.~5 give very similar solutions. Weighted least squares gives a slope of $(6.1 \pm 2.5)\times 10^{-5}$ dex K$^{-1}$, and unweighted least squares gives $(6.9 \pm 2.7)\times 10^{-5}$ dex K$^{-1}$; the $y$-intercept is about $-0.09$ dex. Thus, the slope is significant at about the $2.5\sigma$ level. If we simply compare the number of positive and negative abundance differences for the low and high values of T$_{\rm c}$, we also find support for a positive slope. For example, the differences are negative for 6 elements, positive for 2  elements and zero for one element for T$_{\rm c} < 1200$ K; the differences are negative for 12 elements, positive for 17 elements and zero for 5 elements for T$_{\rm c} > 1200$ K.

Applying the same analysis to the data in Fig.~6, we obtain a weighted least squares slope of $(8.2 \pm 2.7)\times 10^{-5}$ dex K$^{-1}$; unweighted least squares gives $(6.0 \pm 4.4)\times 10^{-5}$ dex K$^{-1}$; the $y$-intercept is about $-0.10$ dex for the weighted fit and $-0.05$ dex for the unweighted fit. It is obvious from Fig.~6 why the weighted solution is different from the unweighted one; three elements (Rb, W and Au) deviate markedly from zero difference, and they have large uncertainties. None of these elements were included in the \citet{lod03} study, because their photospheric uncertainties were too large. Thus, the weighted least squares solution is preferred in this case. For the \citet{asp05} data, the differences are negative for 8 elements and positive for 4  elements for T$_{\rm c} < 1200$ K; the differences are negative for 6 elements, positive for 27 elements and zero for 5 elements for T$_{\rm c} > 1200$ K.

\section{Discussion}

Elemental abundances have been found to correlate with T$_{\rm c}$ in the atmospheres of a few types of stars. \citet{gi05} showed that elemental abundances correlate with T$_{\rm c}$ in the atmospheres of many RV Tauri variables; \citet{gw96} showed that the Type II Cepheid ST Pup shows a similar trend. \citet{lg03} and \citet{a05} argued that the abundance pattern in the A star J37 in NGC 6633 is best explained by accretion of material with high T$_{\rm c}$ elements.

Our analysis of the SWPs' T$_{\rm c}$ slopes above did not reveal significant differences relative to the comparison stars, either as a general trend or as individual outliers. Two of the stars originally flagged by Smith et al. as having high abundance-T$_{\rm c}$ slopes, HD 52265 and HD 217107, were found to be normal.

Our examination of the Solar System abundances revealed a trend with T$_{\rm c}$ in two recent datasets. The simplest interpretation of this trend, assuming it is real, is that the sun's photosphere has been enriched with high T$_{\rm c}$ elements. The quality and quantity of the data plotted in Figs.~5 and 6 are not sufficient to determine if the linear trend can be extrapolated below T$_{\rm c} = 600$ K. If it can be extrapolated to the most volatile elements, then we estimate that elements with the highest T$_{\rm c}$ were enriched in the sun's photosphere by about 0.10 dex; most elements would be enriched by about 0.07 dex. If, instead, the abundance differences between chondrites and the solar photosphere are constant for T$_{\rm c} < 1200$ K, then this implies that high T$_{\rm c}$ elements have been enriched in the sun's photosphere by about 0.07 dex. Thus, the sun's interior metallicity would be smaller than the surface metallicity by about 0.07 dex. The C, N and O abundances in the sun's photosphere should be representative of the primordial Solar System abundances.

While the simplest interpretation of the trends in Figs.~5 and 6 is that the sun's photosphere has been enriched in high T$_{\rm c}$ elements, it is also possible that elements with T$_{\rm c}$ near 600 K have been depleted in the chondrites. Certainly, for elements with T$_{\rm c} < 200$ K (e.g. C, N and O) the depletion in chondrites is very large. Is it safe to assume that elements with $600 \le$ T$_{\rm c} \le 1200$ K are not depleted in chondrites to within the measurement errors?

To answer this, we need to determine if elements with T$_{\rm c}$ just below 600 K are depleted in chondrites. \citet{g03} (footnote 15) noted that the photospheric abundance of In is 0.74 dex larger than the chondritic abundance and cited this as possible evidence for large depletion of In in the chondrites. At the time, however, the published T$_{\rm c}$ estimate for In was 470 K and that of Cd was 430 K, but the Cd chondritic and photospheric abundances are similar. If In is depleted in chondrites, then Cd should be as well, given its lower value of T$_{\rm c}$. This discrepancy seemed to disprove the depletion hypothesis for In and leave its high photospheric abundance unexplained.

\begin{figure}
  \includegraphics[width=3.3in]{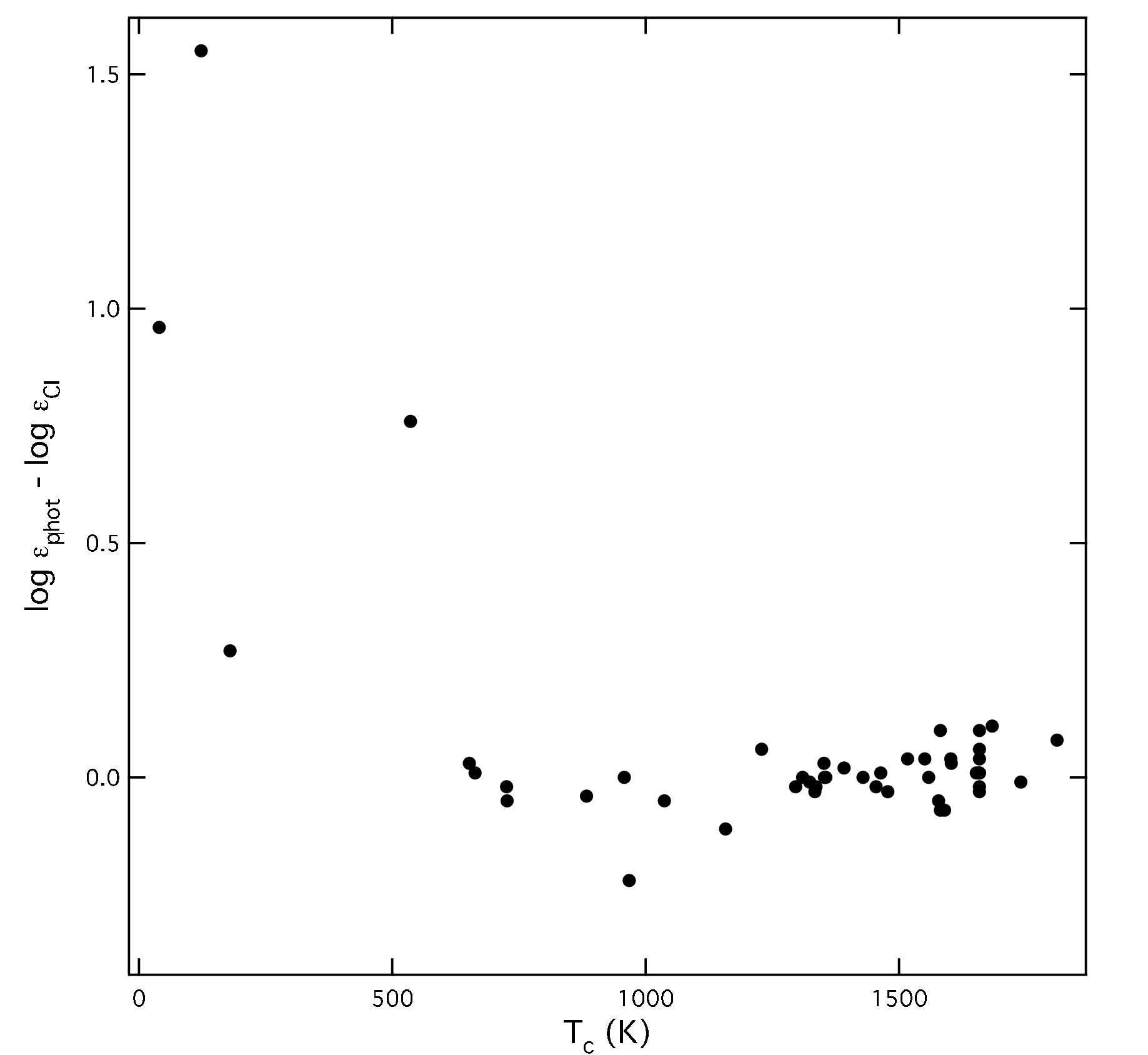}
 \caption{Same as Fig.~5 but also including elements with T$_{\rm c} < 600$ K. In is represented by the point at 536 K.}
\end{figure}

The new T$_{\rm c}$ values published by \citet{lod03} appear to resolve this discrepancy. They are 536 K for In and 652 K for Cd. \citet{lod03} lists only a few other elements with T$_{\rm c}$ values between 200 and 600 K, and only In has a reliable photospheric abundance determination in this range. Br, Hg and Tl have T$_{\rm c}$ values of 546, 252 and 532 K, respectively. Of these, only Tl has a photospheric abundance determination. 

\citet{l72} quote a photospheric Tl abundance range of 0.72 to 1.1 from a sunspot spectrum; this compares to a chondritic abundance of 0.81. We looked for the two Tl I features at 3775 and 5350 \AA\ in the Kurucz Solar Flux Atlas \citep{k84}, but we could only set an upper limit on the Tl abundance near 1.2. It would be worthwhile to revisit the solar Tl abundance with a new umbral spectrum.

We plot in Fig.~7 all the elements from Fig.~5 as well as those elements with T$_{\rm c}$ values $< 600$ K with reliable photospheric abundances. It would be helpful if the abundance differences below T$_{\rm c} = 600$ K displayed a simple trend with T$_{\rm c}$. This would allow us to estimate the amount of depletion in chondrites for elements with higher T$_{\rm c}$. Unfortunately, the few elements in Fig.~7 with T$_{\rm c}$ values below 600 K do not display a simple trend, so we  cannot easily extrapolate to values of T$_{\rm c}$ just above 600 K. Nevertheless, given that In is significantly depleted in chondrites, it seems likely that elements with T$_{\rm c}$ just above 600 K are depleted in by at least a few hundredths of a dex.

\section{Conclusions}

Previous searches for anomalous correlations between photospheric elemental abundances and T$_{\rm c}$ values among SWPs have suffered from either small or heterogeneous samples. We have revisited this topic using a homogeneous set of published abundance data. We fail to find any significant anomalies with respect to T$_{\rm c}$ among a sample of 64 SWPs and 33 comparison stars. 

Nevertheless, we encourage others to revisit this topic as more SWPs are discovered and subjected to detailed spectroscopic analyses. The signature of accretion may reveal itself either as a general trend of abundance-T$_{\rm c}$ slopes with T$_{\rm eff}$ or as anomalously high abundance-T$_{\rm c}$ slopes in a few hotter stars. It would also be helpful to extend the comparison stars sample above [Fe/H] $= 0.2$.

Relative to the chondritic abundances, the solar photospheric abundances do display a weak trend with T$_{\rm c}$. It is seen in two recent compilations of Solar Systems abundances. The implications of this are far ranging, and we encourage additional work on improving photospheric abundance determinations.

\section*{Acknowledgments}

I thank the anonymous referee for very useful comments that improved the paper.

\bsp

\label{lastpage}

\end{document}